\documentclass[10pt,twocolumn, compsoc]{IEEEtran}

\usepackage[hidelinks]{hyperref}
\usepackage{wrapfig}
\usepackage{multirow}
\usepackage{color, colortbl}
\usepackage{enumitem}
\usepackage{balance}
\usepackage[skins]{tcolorbox}
\usepackage{dblfloatfix}
\usepackage{amsmath}
\usepackage{xcolor, graphicx}
\usepackage{algpseudocode}
\usepackage{algorithm}
\usepackage{subcaption}
\usepackage{adjustbox}
\usepackage{tabularx}
\usepackage{lscape}
\usepackage{rotating}
\usepackage{amssymb}
\usepackage{pifont}

\newcommand{\bi}{\begin{itemize}}
\newcommand{\ei}{\end{itemize}}

\newcommand{\revised}{\textcolor{black}}

\begin{document}

\title{  On the Benefits of 
Semi-Supervised Test Case Generation for Simulation Models }
 
\author{Xiao~Ling,
        and~Tim~Menzies,~\IEEEmembership{Fellow,~IEEE}
\IEEEcompsocitemizethanks{\IEEEcompsocthanksitem X. Ling, and T. Menzies are with the Department of Computer Science, North Carolina State University, Raleigh, USA.
\protect 
E-mail: lingxiaohzsz3ban@gmail.com,  timm@ieee.org}}
\markboth{IEEE Transactions on Software Engineering}%
{Ling \MarkLowerCase{\textit{et al.}}: On the Benefits of Semi-Supervised Test Case Generation for Cyber-Physical Systems for IEEE Journals}

\IEEEtitleabstractindextext{
\begin{abstract}
Testing complex \revised{simulation models} can be expensive and time consuming.
Current state-of-the-art methods that explore this problem are fully-supervised; i.e. they require that all examples are labeled.
On the other hand, the GenClu system (introduced in this paper) takes a semi-supervised approach;
i.e. (a)~only a small subset of   information is actually labeled (via simulation) and (b)~those labels are then spread
across the rest of the data.  
When applied to five open-source \revised{simulation models of
cyber-physical systems}, GenClu's test generation can be multiple orders of magnitude faster
than the prior state of the art.
Further, when assessed via mutation testing, tests generated by GenClu were as good or better than anything else tested here. 
Hence, we recommend semi-supervised
methods over prior methods (evolutionary search  and 
fully-supervised learning).


To enable open science, all the data and scripts used in this study are available at \url{https://github.com/ai-se/CPS_test_generation}. 
\end{abstract}


}
\maketitle
 
\begin{figure*}[!t]
    \centering
    \includegraphics[width=0.9\textwidth]{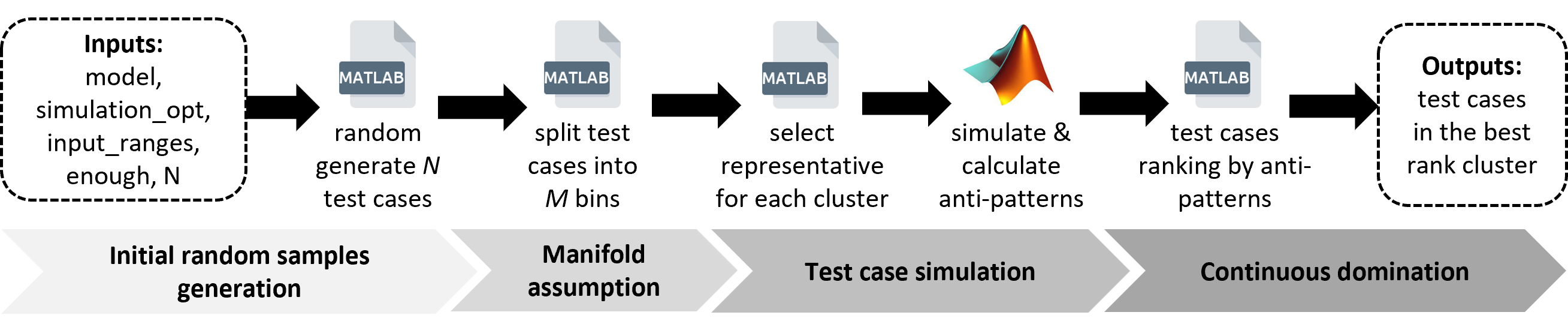}
    \caption{Framework of our proposed clustering test case generation approach GenClu. As we can see from this figure, the clustering method is a straight forward method which does not need any iteration or evolution.}
    \label{fig:clustering}
\end{figure*}

\section{Introduction}\label{Introduction}

Simulation-based testing can be very slow and expensive. This is especially true for (e.g.)   complex cyber-physical systems (CPSs) that combine software with physical \revised{components}
(since each such simulation can take several times longer than traditional software~\cite{arrieta2016search}).   Arrieta et al. warn that  testing a complex system requires hours to days of CPU~\cite{arrieta2019pareto, gonzalez2018enabling, sagardui2017multiplex}. 
\revised{
Traditional approaches for \revised{simulation model test} generation assume that all examples can be labeled. \revised{Since generating those labels may require simulation, the labeling process can be very slow. For example, EPIcuRus is ``fully supervised’’ method, so its algorithm needs labels for every example.} Other approaches use even slower meta-heuristics algorithm (e.g genetic algorithms)~\cite{arrieta2017search, arrieta2019pareto, arrieta2016search, matinnejad2018test}.}

\revised{
To address the labeling problem, we propose a semi-supervised ``label-lite'' method called GenClu (that needs very few labels) based on unsupervised clustering. As shown in Figure~\ref{fig:clustering},  GenClu:} generates test cases randomly. This is done by picking values at random from the minimum to the maximum of known values (note that this process generates valid test case inputs, for which the test case output is unknown).

Next, based on the manifold assumption (described in \S\ref{se}), GenClu \revised{groups} a large number of candidates into small clusters based  on their independent attributes. For this clustering, we use a recursive FASTMAP clustering algorithm~\cite{faloutsos1995fastmap, platt2005fastmap}. 
Then, we can find the best group of candidates by simulating only one representative from each cluster. 

An interesting feature of GenClu is that there is no need to go further and build another model (e.g. via decision tree classification) since, for GenClu, {\em the clusters are the model}. This means that in theory, GenClu is much faster than \revised{meta-heuristic search or fully-supervised approaches} since  (a)~it needs far fewer runs of the simulator, and (b)~it has no need for multiple modeling steps. The experiments in this paper confirm these theoretical predictions:
\bi
\item When assessed via mutation testing, GenClu is as good or better than anything else tested here \revised{(RQ1)}.
\item \revised{GenClu runs magnitude faster than OD and SAMOTA system.}
In practice, these speeds mean that GenClu takes minutes to terminate while other methods require hours \revised{(RQ2)}. 
\item \revised{We find that, with the tree structure generated by GenClu, we can mutate the test cases in the first few best clusters and generate new test cases that have improved performance compared to other clusters (RQ3).}
\ei
Based on the above, for generating test cases for simulation models, we would recommend semi-supervised method over evolutionary search (as done in OD), \revised{or the surrogate model based optimization approach (as done in SAMOTA)}.

\revised{The rest of this paper is structured as follow: Section~\ref{background} presents the necessary and important background of this study, as well as the related work. Section~\ref{clusterapproach} illustrates our proposed GenClu framework. Section~\ref{methodology} briefly introduces two state-of-the-art algorithms OD and SAMOTA. Section~\ref{experiment} illustrate the case studies, mutants generation, performance criteria, and the statistical analysis. Section~\ref{result} presents our experimental results and the conclusions to RQs based on the results. Section~\ref{threats_to_validity} explains the threats that may have in this study, and Section~\ref{conclusion} summaries this study and proposes some future works.}

\section{Background}\label{background}
\subsection{Semi-Supervised Learning}\label{se}


Semi-supervised learners (SSL) seek to ``find the best model from a small number of data points which have labels''~\cite{tu2021frugal}. For example, the {\em active learner} is a kind of SSL learner which (a) only labels a group of representative data points, (b) builds a model based on those data, and (c) uses that model to find the next group of representative data points that need to be labeled~\cite{kocaguneli2012active}. SSL has been widely applied in traditional SE tasks (e.g. test case prioritization~\cite{yu2019terminator}, finding relevant literature~\cite{yu2018finding}, and software analytics~\cite{tu2021frugal}) which label all data points need a high budget. 

This paper asks if clustering based SSL techniques can ``select'' the most relevant test cases in a randomly generated test set\footnote{\revised{To be precise, our system starts with some randomly generate tests, from which we select a small set of the most informative tests. One way to view this is ``information condensation''; i.e. initially, we have no knowledge about the value of each test, but after clustering, we have grouped together similar tests, thus gathering enough testing ``condensate'' in one place to make an informed selection about what tests are useful.}}.
This is a reasonable question since testing on the simulation model can be extremely time consuming. For example, high fidelity systems (e.g. drone, elevator, and AV) can take several hours to a few days to simulate a task~\cite{arrieta2016search, arrieta2017employing, arrieta2019pareto}. Thus, labeling all test cases (e.g. 100 to 150 test cases) can take an inordinately large amount of  time. This motivates us to find a test case generation approach that requires fewer labels, and hence, \revised{a smaller number} of simulations.  

SSL is a technique that makes a {\em manifold assumption} about the data explored by a data miner:
\bi 
\item
Many high-dimensional data sets that occur in the real world actually lie along low-dimensional latent manifolds inside that high-dimensional space~\cite{cayton2005algorithms}.
\item
Hence, domains that are seen to use many attributes can actually be described by a comparatively small number of attributes without loss of signal.
\ei
The manifold assumption has implications for how we explore the world. The space of options in any data set is the cross product of the cardinality $C$ of each attribute $A$; i.e. $C^A$. According to the manifold assumption, the effective number of attributes is $A' \subset A$. Hence, the space of options is actually $C^{A'}$, which, when
$|A'|\ll |A|$, is a much smaller space to explore than looking at all values of all attributes.

One way to view SSL is as an intelligent sampling strategy to learn the most from the fewest observations.
{\em Sampling} is a statistical method that selects informative samples from the large initial \revised{set of} samples to carry out the experiment. Kiaer et al.~\cite{kiaer1895observations} are the leading researchers who used sampling methods in social and economic data in 1895. Several famous probability sampling techniques make extensive use of clustering, including random sampling, systematic sampling, stratified random sampling, clustering sampling, and multistage sampling~\cite{taherdoost2016sampling}.  
Here, we extend all these sampling methods with a Nystr{\"o}m algorithm (an approximation of the eigenvectors of a large matrix, based only on a rectangular submatrix of the large matrix~\cite{platt2005fastmap}). By clustering on the eigenvectors, GenClu can find good clusters that most divide the data according to its ``shape'' (as defined by the eigenvectors). For more details on our particular Nystr{\"o}m approximation, see the clustering methods of \S\ref{clusterapproach}.



As our best knowledge, this study is the first study to utilize sampling via Nystr{\"o}m  clustering  to the task of testing Cyber-Physical Systems. 


\subsection{Cyber-Physical Systems (CPSs)}\label{simulation}
CPS systems are embodied in their environment and   at least in our case studies, these systems make extensive use of \revised{control theory~\cite{aastrom2021feedback}}. In that theory, the feedback controller is used to compare the value or status of process variables with the desired set-point. This controller then applies the difference as a control signal to bring the process variable output of the plant to the same value as the set-point~\cite{ling2021faster}.
Hence, many CPS developers make extensive use of simulation-based testing.
This approach is widely used since high-fidelity simulators are often generated as part of the CPS construction~\cite{matinnejad2016automated}.

\begin{figure}
    \centering
    \includegraphics[width=0.48\textwidth]{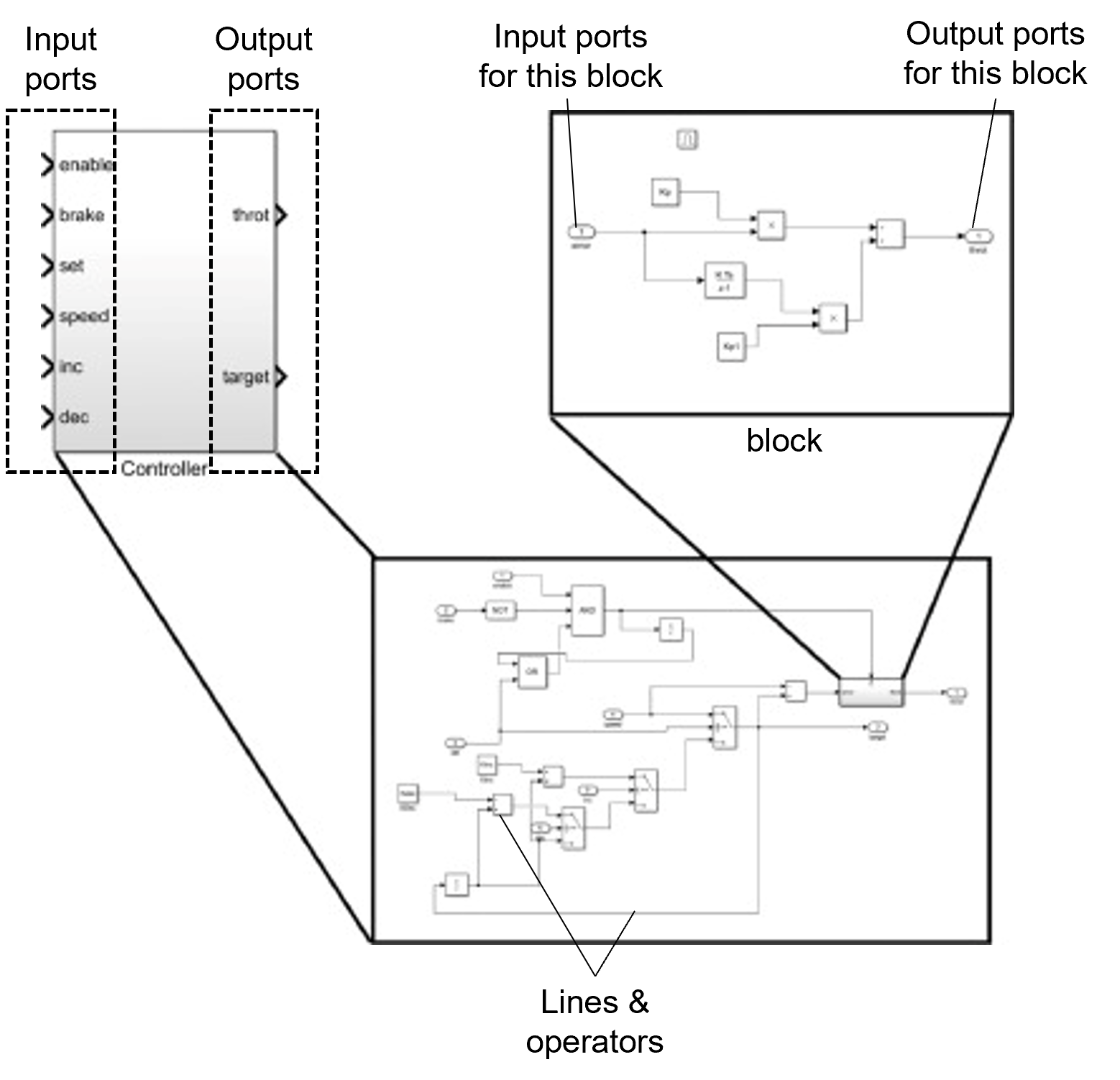}
    \caption{An example of a Simulink model - Cruise Controller of a car~\cite{arrieta2019pareto, ling2021faster}.}
    \label{fig:simulink}
\end{figure}

Developers of CPSs usually rely on some simulation development tools (e.g. Simulink) to build the system~\cite{chowdhury2018automatically}. Figure~\ref{fig:simulink} shows an example Simulink model (i.e. Cruise Controller of a car) with only two hierarchical levels~\cite{arrieta2019pareto, ling2021faster}. As we can see, the overall system is contained in a container with 6 input ports and 2 output ports. In that container, we can find that the whole system is combined with lots of logical and arithmetic operators, switch blocks, and relation operators, which are connected by the lines. Also, the sub-block performs like a ``function'' in a normal coding manner which contains more operations for a certain purpose. Moreover, the sub-block has its own input ports and output ports which perform like the input and output of a ``function''.

Very different to the normal software, the inputs $x = \{ x_1, x_2, \cdots, x_n \}$ and outputs $y = \{ y_1, y_2, \cdots, y_m \}$ of simulation models are signals (here signal means the time series vector). More specifically, simulation models simulate the results in a duration of time domain $T = [0, b]$, where each time step has its corresponding input and output value. \revised{The delta between two time stamps is called time step} ($\Delta T$) and is usually defined by developers. For example, one simulation model has simulation time set as 10 seconds ($T = [0, 10]$), and the time stamp is 0.001, then there will be $10/0.001 + 1 = 10001$ simulation steps. This can indicate that the inputs and outputs of this model should be a vector of length 10001.

\subsection{Testing Simulation Models}\label{testingsimulation}

Test case generation for simulation models has been widely studied in the past decades. The common goal for test case generation is that generating \textit{efficient} test suite (i.e. we can run fewer tests) while increasing the \textit{effectiveness}~\cite{wong1997test, wong1998effect, ahmed2016test, di2015coverage, yoo2012regression}. However, finding a balance point between these two goals is always a challenge.

Ling et al~\cite{ling2021faster} discusses features of \revised{simulation model} testing that make it different from testing other
kinds of software. simulation models (especially the CPSs) are embodied in their environment, so only exploring static features (e.g. the code base) as traditional software does is not enough. Testing on these models \revised{requires} to test how that code base reacts to its surrounding environments. Thus only exploring static features is not recommended for testing these models. 

Further, \revised{simulation models} make extensive use of process control theory, where a feedback controller is used to compare the value or status of process variables with the desired set-point. Moreover, running these models is sometimes cost expensive since lots of models require one to few days for a single execution~\cite{arrieta2017search, arrieta2019pareto, matinnejad2018test}. Thus collecting huge test case execution data or coverage information is a hard and expensive job for simulation models. \revised{Hence, the data obtained from the simulation models (e.g. input vectors and output vectors) is the only source for testing the system}. In this study, we only use the input and output signals in the simulation models instead of test execution history or coverage information to generate an adequate test suite.

Our goal for this study is to generate an adequate test suite for simulation models which can detect as many faults as possible. \revised{As we will talk later, the ``faults'' in our study are artificial added mutants, which we use to judge the performance of test cases}. When tracing the performance goal, we prefer a test suite that contains fewer test cases in it. Here, we can define the test case generation problem as follow:
\begin{tcolorbox}[boxsep=1pt,left=4pt,right=4pt,top=2pt,bottom=2pt]\small
    Given a simulation model $M$, and define an evaluation function $f$ to evaluate the performance of test case generation approach, we want to find a test suite generation approach $P_{opt}$ such that the test suite $TS_{opt}$ generated by $P_{opt}$ has the highest performance. More specifically, $f(P_{opt}) > f(P)$ and $1 \leq |TS_{opt}| \leq |TS|$.
\end{tcolorbox}

Gaaloul et al. stated the test generation challenges in the simulation model as follow~\cite{gaaloul2020mining}:
\begin{itemize}
    \item Test cases generated for simulation models should be in the signal format (functions over time) in most of the cases unless the inputs are constants.
    \item Generated signals must be meaningful for their corresponding models and should be realistically generated. \revised{Here the ``meaningful'' means the values cannot be very large or very small which the simulated model cannot achieve in the real world.}
\end{itemize}

\begin{figure}[t]
    \centering
    \begin{subfigure}[b]{0.48\textwidth}
       \includegraphics[width=1\linewidth]{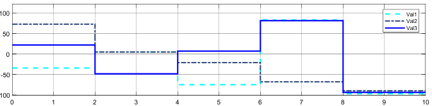}
       \caption{Input signals}
       \label{fig:input} 
    \end{subfigure}
    
    \begin{subfigure}[b]{0.48\textwidth}
       \includegraphics[width=1\linewidth]{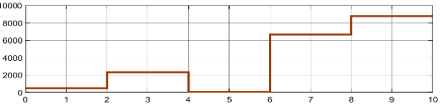}
       \caption{Output signal}
       \label{fig:output}
    \end{subfigure}
\caption{An example of input signals and output signals}
\label{fig:example}
\end{figure}

\begin{table*}[!t]
    \centering
    \begin{tabular}{c|c|c|c|c}
        {\bf Name} & {\bf Description} & {\bf Goal} & {\bf Formula} & {\bf Example}\\
        \hline
        \multirow{8}{1.6cm}{Discontinuity} & \multirow{8}{2cm}{Short duration pulse in the output signal~\cite{matinnejad2017automated}} &  \multirow{8}{*}{Max} & \multirow{4}{*}{$dis(O_i) = \max\limits_{dt=1}^{3}(\max\limits_{j=dt}^{k-dt}(min(lc_j, rc_j)))$} & \multirow{8}{*}{\includegraphics[width=0.14\linewidth]{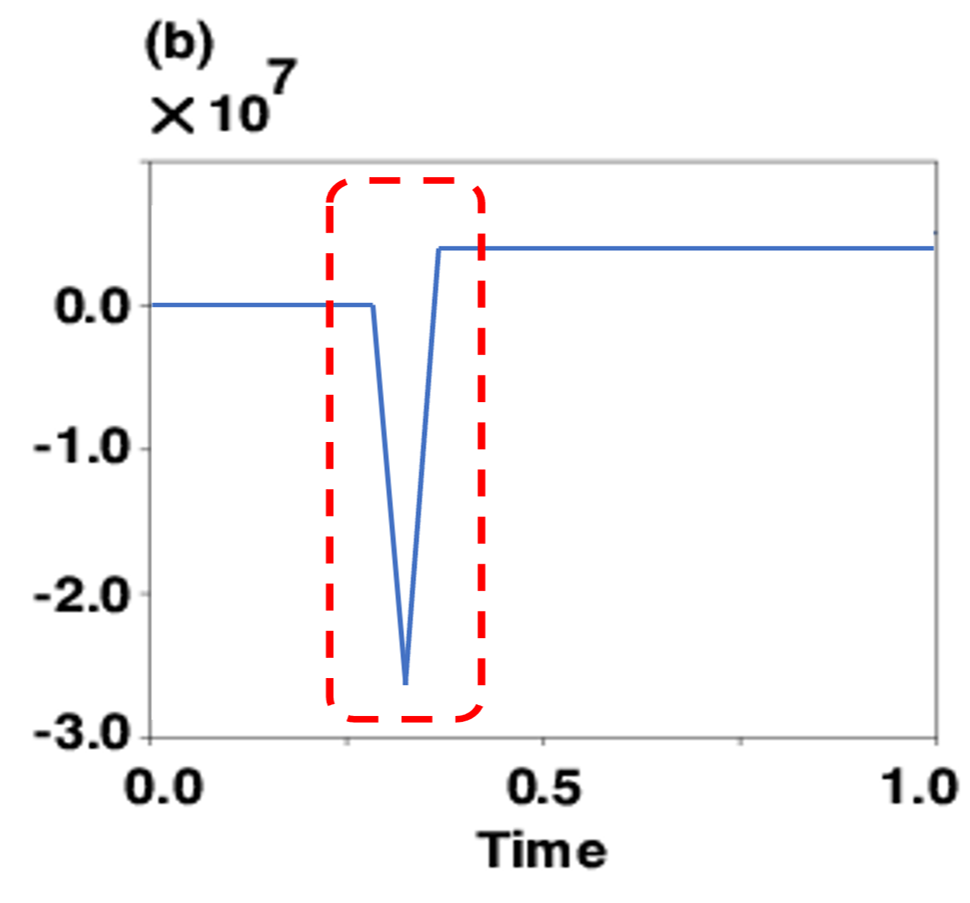}} \\
         & & & & \\
         & & & & \\
         & & & & \\
         & & & $lc_j$ (left change rate) $= |sig(j \cdot \Delta t) - sig((j-dt) \cdot \Delta t)| / \Delta t$ & \\
         & & & $rc_j$ (right change rate) $= |sig((j+dt) \cdot \Delta t) - sig(j \cdot \Delta t)| / \Delta t$ & \\
         & & & & \\
         & & & & \\
        \hline
        \multirow{8}{1.6cm}{Instability} & \multirow{8}{2cm}{The duration of quick and frequent oscillations in the output signal~\cite{matinnejad2017automated}} & \multirow{8}{*}{Max} & \multirow{8}{*}{$ins(O_i) = \sum_{j=1}^{k}|sig(j \cdot \Delta t) - sig((j-1) \cdot \Delta t)|$} & \multirow{8}{*}{\includegraphics[width=0.14\linewidth]{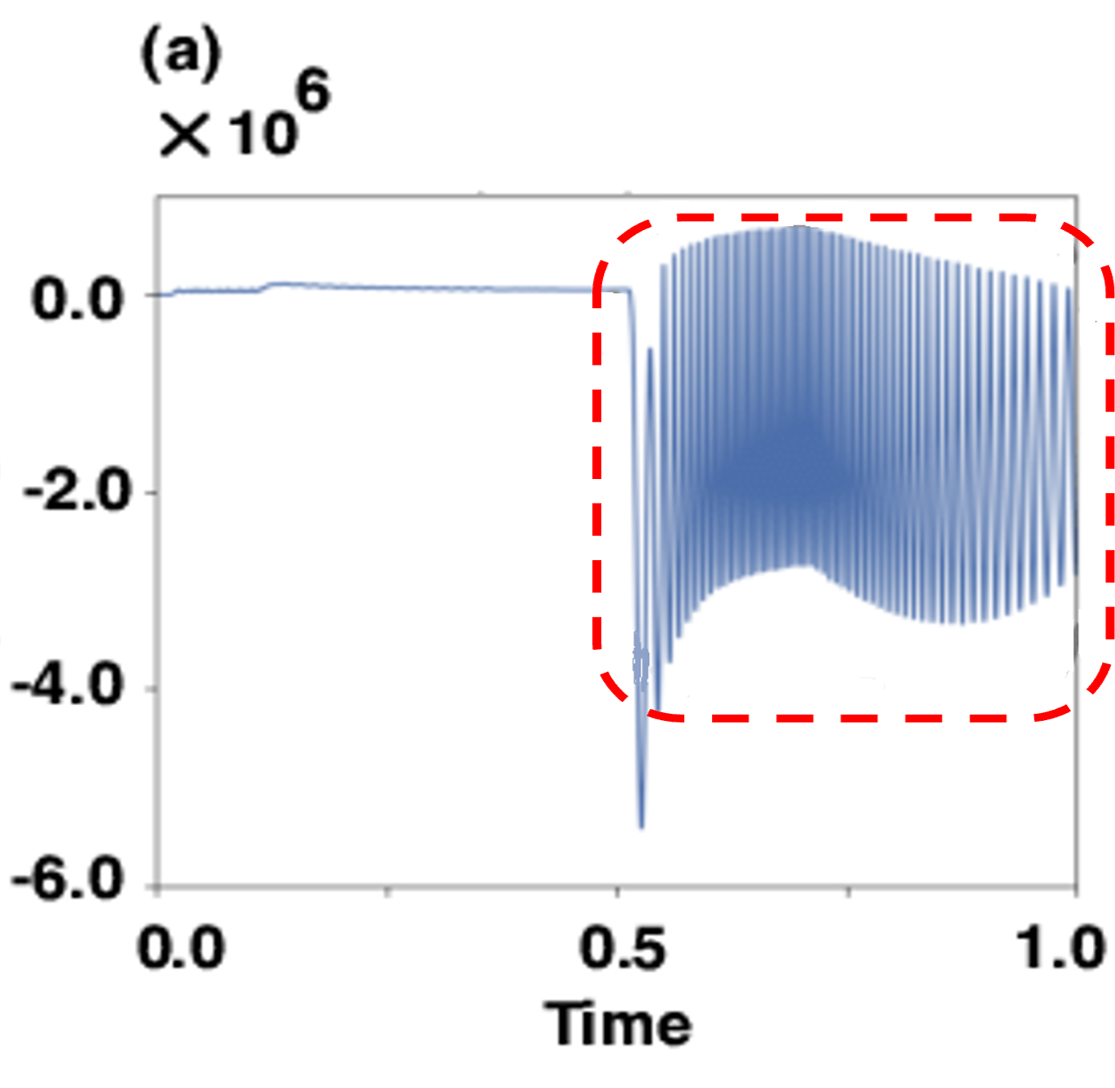}} \\
         & & & & \\
         & & & & \\
         & & & & \\
         & & & & \\
         & & & & \\
         & & & & \\
         & & & & \\
        \hline
        \multirow{8}{1.6cm}{Growth to Infinity} & \multirow{8}{2cm}{Output signal increased or decreased to an infinity value~\cite{matinnejad2015effective}} & \multirow{8}{*}{Max} & \multirow{8}{*}{$inf(O_i) = \max\limits_{j=1}^{k} |sig(j \cdot \Delta t)|$} & \multirow{8}{*}{\includegraphics[width=0.14\linewidth]{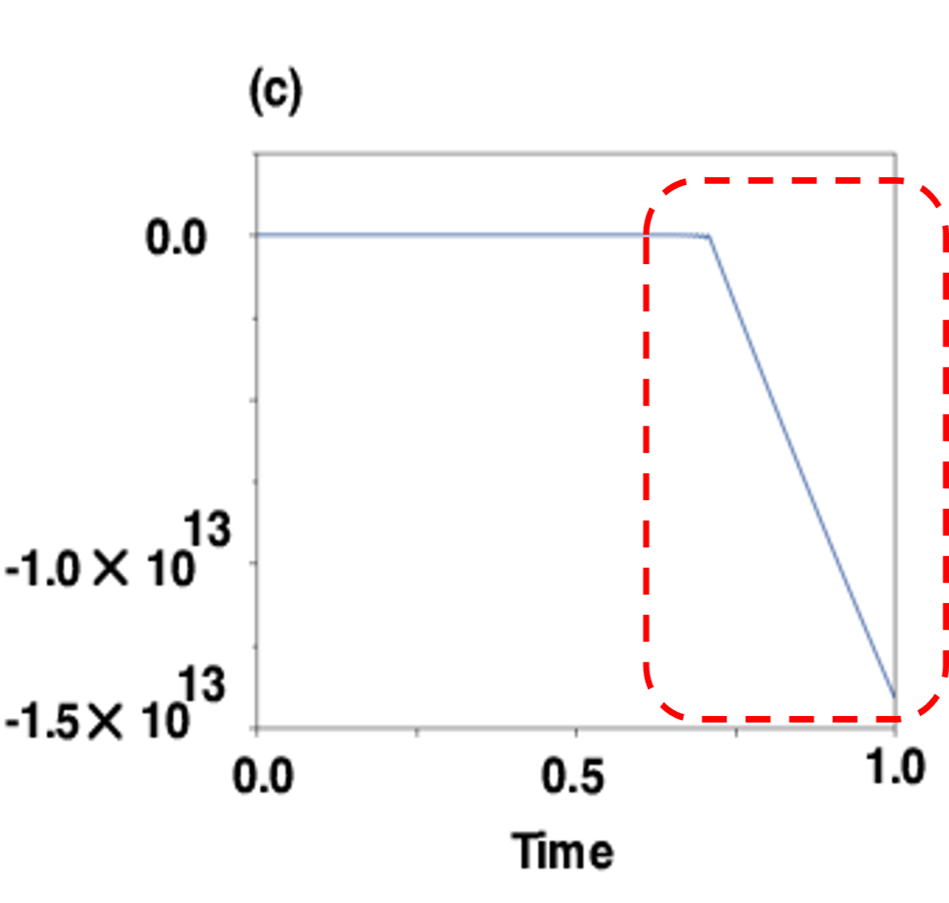}}\\
         & & & & \\
         & & & & \\
         & & & & \\
         & & & & \\
         & & & & \\
         & & & & \\
         & & & & \\
        \hline
        \multirow{8}{1.6cm}{MinMax} & \multirow{8}{2cm}{How much the model is detected by the test case~\cite{arrieta2019pareto}} & \multirow{8}{*}{Max} & \multirow{8}{*}{$minmax(O_i) = |\max\limits_{j=1}^{k}(sig(j \cdot \Delta t)) - \min\limits_{j=1}^{k}(sig(j \cdot \Delta t))|$} & \multirow{8}{*}{~} \\
         & & & & \\
         & & & & \\
         & & & & \\
         & & & & \\
         & & & & \\
         & & & & \\
         & & & & \\
    \end{tabular}
    \caption{Collection of anti-patterns for simulation output signal from literature~\cite{arrieta2019pareto, matinnejad2015effective, matinnejad2017automated}. In formulation, $O_i$ means the $i^{th}$ output, $k$ means the total number of simulation steps, and $\Delta t$ represents the time stamp in simulation.}
    \label{tab:anti-pattern}
\end{table*}

To address these two challenges, Gaaloul et al.~\cite{gaaloul2020mining} used the approach of Matlab that generate signals by using (a) interpolation function ($i_u$), (b) input domain ($r_u$), and (c) number of control points ($c_u$). Here the control points represent the special points in the input signals where the signal pattern changes. In this study, we adopt this approach and follow the generation patterns in their work. More specifically, we create a Matlab code which takes the above three requirements as inputs, and returns $n$ signals where $n$ is the number of inputs. For each control point $c_{u_k}$ in $m$'s input signal $I_m$, the code randomly generates a value in the range $r_{u_m}$ and makes control points equally distributed over the time domain $T$. After that, the interpolation function is utilized to generate a signal that connects the control points. Gaaloul et al. stated that the interpolation function Matlab included are others (user defined), linear, piecewise constant and piecewise cubic interpolations~\cite{gaaloul2020mining}. In our study, all models use piecewise constant interpolation as we checked from their literature except one model which all inputs are constant. Figure~\ref{fig:example} shows an example of input signals generated by our code and the corresponding output signals. As we can see from Figure~\ref{fig:example}(a), there are five control points represent the values of signals at time 0, 2, 4, 6, and 8 respectively. All values are random generated in the input ranges $\{[-100, 100]; [-100, 100]; [-100, 100]\}$.

\subsection{The Oracle Problem in Simulation Models}\label{blackboxData}
One way to characterize this work is that we study how to test in simulation models when humans may not know all the possible nominal and off-nominal behaviors of a system. This is called the {\em oracle problem} and is particularly acute in \revised{simulation models}~\cite{arrieta2016search, arrieta2019pareto, matinnejad2018test}.
We address the oracle problem using two techniques:
\bi
\item Anti-patterns especially developed for \revised{simulation models} (see \S\ref{ap});
\item Mutation testing (see \S\ref{mutationtesting}).
\ei

\subsubsection{Simulation Model Anti-Patterns}\label{ap}

Simulation output signals \revised{can} have some undesirable patterns which can indicate the faulty behavior during the simulation. Those patterns are called anti-patterns and have been widely studied in some past literature~\cite{ling2021faster, arrieta2019pareto, wang2013minimizing, matinnejad2017automated, matinnejad2015effective} where \revised{a human oracle is not available}. We also utilize those anti-patterns in our experiments to address the oracle problem.

By looking through the past literature, we have collected four anti-patterns that related to our study and can be obtained with our settings. They are {\em discontinuity}, {\em instability}, {\em growth to infinity}, and {\em minimum \& maximum difference}. \revised{Note that the {\em minimum \& maximum difference} is more about the effectiveness metric instead of the anti-pattern. However, to simplify the description of those metrics, we include it in the illustration of anti-patterns.} As stated above, those terms have shown the success in revealing faulty behaviors in the past literature. We have summarized those terms in Table~\ref{tab:anti-pattern}:
\bi{\item The }{\bf description} column summarizes each anti-pattern;
\item The {\bf goal} column shows how these anti-patterns can most likely reveal faults from output signals, 
\item The {\bf formula} column presents detailed mathematical formulas on how to calculate those anti-patterns.
\item The {\bf example} column presents the examples of anti-patterns in the signal format.
\ei
Note that all our   output signals will have 4 anti-pattern values. Therefore, in our proposed clustering approach, the total number of goals will be $4 * n$ where $n$ represents number of outputs in a certain \revised{simulation} models. We use the
\textit{domination}  predicate (described below) to select better candidates on these $4 * n$ goals.

In multi-goal optimization, {\em domination predicates} assess how well an optimizer is trading off between multiple goals. Traditionally, this is done with {\em binary domination}, which states that one solution is better than another if none of the goal is worse than the others and at least one solution is better\footnote{To say that another way, we have {\em zero reasons} to prefer the other solution and at least {\em one reason} to prefer this solution.} 

Many studies warn that binary domination struggles to distinguish candidates for three or more goals~\cite{Zitzler2004,sayyad2013value,Wagner07}. Hence Zitzler~\cite{Zitzler2004} proposed a {\em continuous domination predicate} that reflects on the {\em size} of the goal value delta ($\Delta(a,b)$)  between two solutions $a,b$. Given $1 \le i \le n$ goals and some weight value $w_i$ that is $(1,-1)$ for goals we want to minimize or maximize, then Zitzler says that moving from solution $a$ to solution $b$ will ``cost'' \mbox{$\Delta(a,b)=\sum_i  e^{w_i * (a_i-b_i)/n}$}. He then concludes that solution $a$ is better than solution $b$ if it loses least; i.e. \mbox{$\Delta(a,b) < \Delta(b,a)$}.

\subsubsection{Mutation Testing for Simulation Models}\label{mutationtesting}
Another way to address the oracle problem is the  use of {\em mutation testing} to evaluate the performance of different test case generation approaches. Mutation testing is a fault-based software testing technique~\cite{jia2010analysis, papadakis2019mutation, pizzoleto2019systematic,schuler2009javalanche,offutt1996experimental}. It checks the ability of test cases to reveal some artificial defects~\cite{papadakis2019mutation}. The defects or faults generated from the original software are called mutants. In traditional software, mutants are created by systematically injecting small artificial faults into the program under test~\cite{just2014mutants}. However, mutants generation for Simulink models is quite different to the traditional software. Simulink models have their certain fault patterns (which we will introduce in Section~\S\ref{mutation}) and the action of seeding artificial faults may include (a) changing arithmetic operators, (b) changing constant values, and (c) changing relation operators, etc~\cite{matinnejad2018test}. Due to certain fault patterns, the mutants for Simulink models may not be as many as traditional software has (As our observation, usually around 10-100 on the open-source simulation model after mutant preprocessing). Even the number of mutants is low in Simulink models, detecting all of them still be a challenge (For example, in this study, only 10-15\% of repeated experiments get full mutation score on the small test suite). In Section~\S\ref{mutation}, we will detailed introduce (a) fault patterns in Simulink models, (b) how to manually generate mutants for our case studies, and (c) how to preprocess those manually generated mutants.

\subsection{Other Related Work}
For the sake of completeness, this section discusses other work in the area (even if much of the following is not \revised{directly} related to our problem of mutation testing guided by semi-supervised learning for the detection of anti-patterns in the simulation models).
\bi 
\item
Much prior work explores test case generation for different case studies or various requirements. In 2003, Zhao et al. used a genetic algorithm and simulation models of a CPS to created a test case generation approach. They utilized a coverage measure based on the formulation of a coverage problem they defined~\cite{zhao2003generating}.
\item
In 2008, Gadkari et al. utilized temporal logic and translated it into formal language in high-level requirements and test specifications to generate   test cases~\cite{bashir2008automated}. 
\item
In 2013, Matinnejad et al. proposed a search-based test case generator    for continuous controllers~\cite{matinnejad2013automated}.
\item
In 2015, Sinha et al. proposed a framework utilized the high-level functional requirements to automatically generate test cases. In their study, requirements are specified in the natural language and recognized by a formalized ontology. The ontology is refined with controller and plant models during development so that test cases can automatically generated on any stage~\cite{sinha2015requirements}.
\item
In 2016, Matinnejad et al. presented another search-based test case generation  for time-continuous simulation case studies. They firstly relied on engineers to defined a sequence of signal segments, and then selected some  inputs to generate test cases~\cite{matinnejad2016automated}. 
\item
In 2017, Arrieta et al. applied Non-dominated Sorting Genetic Algorithm II on three cost-effectiveness measures (i.e. requirements coverage, test case similarity, and test execution time). They proposed one crossover operator and three mutation operators (at test suite level, test case level, and both level)~\cite{arrieta2017search}.
\item
Also in 2017, Arrieta et al. stated the challenges that simulation model of industrial CPSs are very complex and executing them became computationally expensive. To address these challenges, they proposed a fitness function with four objectives and evaluated five multi-objective search algorithms on their case studies. They found NSGA-II outperformed other multi-objective search algorithms on this task~\cite{arrieta2017employing}.
\item
In 2018, Gonzalez et al. proposed to conduct testing at early stages and over executable models of the system and its environment. Thus they proposed a SysML-based modeling methodology and an effecient Simulink co-simulation framework~\cite{gonzalez2018enabling}. 
\item
Also in 2018, Turlea et al. proposed a discrete genetic test case generation approach which aimed to reveal more requirements violations at the model-in-the-loop test level~\cite{turlea2018search}. 
\item
In 2019, Zhang et al. stated CPSs typically operate in highly indeterminate environment that testing on such environment has lots of uncertainties. Thus they proposed two test case generation based on the Uncertainty Theory and multi-objective search~\cite{zhang2019uncertainty}. 
\item
In 2020, Menghi et al. warned that the compute-intensive CPSs for a large number of test inputs is very time consuming. Hence they proposed ARIsTEO which built the surrogate models by using the inputs and outputs from a single simulation result, and utilized the falsification algorithm to find an input which violated the requirements in those surrogate models~\cite{menghi2020approximation}. 
\ei


Another important related work is the EPIcurUS system~\cite{gaaloul2020mining} which is worthy of its own subsection.

\subsubsection{Open-loop Modeling and Testing with EPIcuRus}\label{epi}
\revised{
Closed-loop models interact with their own components while open-loop modeling explores the  interaction with other systems. All the above systems, including GenClu, assumed to test the closed-loop models, while EPIcuRus~\cite{gaaloul2020mining} tests for the open-loop system. EPIcuRus combines decision tree classifiers (to explore the potential assumptions that can satisfy the requirements) with model checking (to evaluate the assumptions). In each iteration, the promising ranges returned from the decision tree data miner will be used to generate test cases for the next iteration.}

\revised{``Testing'' means something different for EPIcuRus and GenClu. We ``test'' to obtain some scores from an oracle. EPIcuRus ``tests'' to explore the assumption space within and without the model.}

\revised{That said, internally, EPIcuRus and  GenClu share some similar internal structures. Both systems run a learner, then query the output.  GenClu uses semi-supervised learning to recursively cluster the data, then probes the leaf clusters. EPIcuRus uses a fully supervised tree learner, the branches of which are then queried to find the attribute ranges that separate different outcomes (specifically, EPIcuRus' test engine returns the assumptions under which tests may fail).}

\revised{In principle, GenClu could be extended to implement EPIcuRus-like functionality (e.g. if we ran a decision tree learner to find the deltas between our leaf clusters). But that is a matter for future work.}

\section{GenClu - Test Suite Generation Approach with Clustering}\label{clusterapproach} \revised{
In this section, we will introduce our proposed test suite generation framework called GenClu. Different from the other approaches based on evolutionary algorithms and fully supervised approaches that require labeling \textbf{ all} candidates, GenClu only needs to label \textbf{ a few} candidates by (a) clustering the unlabeled examples, (b) using a simulator, labeling a few examples per cluster, and (c) sharing the fitness values around the local cluster.
}

\revised{
Figure~\ref{fig:clustering} shows \revised{the} overall framework of our proposed clustering approach. In stark difference to the methods described in the last section, instead of finding the optimal solution in multiple iterations, GenClu's clustering approach does all its work in just one iteration:
\bi 
\item First, we generate test case candidates, at random, as follows. Given knowledge of the min/max values of each attribute, we pick at random from that range. As suggested by previous literature~\cite{matinnejad2018test}, the signals change their values in some set-points. We generate values in the range in those set-points, and those values will be the features of our clustering algorithms. Note that this process generates valid test case inputs, for which the test case output is unknown.
\item Next, \revised{under} the manifold assumption of \S\ref{se}, GenClu \revised{groups} a large number of candidates into small clusters (based on their independent attributes). For that clustering, we use a recursive FASTMAP clustering algorithm~\cite{faloutsos1995fastmap, platt2005fastmap}. 
\item
Then we can find the best group of candidates by simulating only one representative from each cluster. The output of our algorithms will be test cases that in the best group.
\ei
}
\begin{algorithm}[!b]
    \caption{\textbf{cluster}: Clustering method to build the tree structure}
    \label{alg:cluster}
    \small
    \begin{algorithmic}[1]
        \Require testSuite, enough, tree, node
        \If{\textbf{size}(testSuite) $>$ enough}
        \State eastlist, westlist = \textbf{split}(testSuite) \Comment{two lists of items}
        \State node.left, node.right = eastlist, westlist
        \State \textbf{cluster}(eastlist, enough, tree, left\_node)     \Comment{recuse}
       \State \textbf{cluster}(westlist, enough, tree, right\_node)    \Comment{recuse}
        \EndIf
    \end{algorithmic}
\end{algorithm}
\begin{algorithm}[!b]
    \caption{\textbf{split}: Split candidates into east region and west region by using the FASTMAP technology~\cite{faloutsos1995fastmap, platt2005fastmap}, return eastItems and westItems}
    \label{alg:split}
    \begin{algorithmic}[1]
        \Require testSuite
        \State pivot = \textbf{random}(testSuite) \Comment{pivot is any point} 
        \State east = \textbf{mostDistance}(testSuite, pivot) \Comment{Farthest   to   pivot}
        \State west = \textbf{mostDistance}(testSuite, east) \Comment{Farthest   to   east}
        \State c = \textbf{distance}(east, west) 
        \For{point \textbf{in} testSuite}
            \State a = \textbf{distance}(point, east)
            \State b = \textbf{distance}(point, west)
            \State point.d = (a$^2$ + c$^2$ - b$^2$) / (2c)
        \EndFor
        \State new\_testSuite = \textbf{sort}(testSuite.d) \Comment{Sort all  via  d}
        \State eastItems = new\_testSuite[:0.5*\textbf{size}(testSuite)]
        \State westItems = new\_testSuite[0.5*\textbf{size}(testSuite):]
    \end{algorithmic}
\end{algorithm}
\noindent\revised{
For details on our clustering  see   Algorithm~\ref{alg:cluster}:
\bi 
\item
In the \textbf{cluster} algorithm, line 1 checks if the size of current group of candidates exceeds the value in "enough" parameter. If it is larger than "enough", then we need to cluster these candidates again. \revised{Please note that the ``enough'' parameter in our study means the size of the clusters.}
\item 
Line 2 uses the \textbf{split} algorithm to divide all candidates into two bins. For details of \textbf{split}, see Algorithm~\ref{alg:split}).
\item
Line 3-5 puts two bins into  a tree structure and recursively calls \textbf{cluster} algorithm on the new groups until no more split can be made. In our experiment, we only evaluate items in the most bottom nodes, which no more split can be applied to them.
\ei
As to the \textbf{split} Algorithm~\ref{alg:split}:
\bi 
\item The code first selects one random candidate as pivot (line 1).
\item After that, it finds the most distance point to the pivot as east group representative (line 2). 
\item The most distance point to the east representative is marked as west group representative (line 3). 
\item After that, each candidate in the test suite is assigned to the east group or the west group by using the \textit{cosine rule} (lines 5..8). Here the distance function we used is the \textit{Euclidean distance}. 
\ei
One interesting point of this approach is that data is clustered without comparing all points.
Hence, instead of requiring $O(N^2)$ distance calculations, this algorithm terminates in the time needed for the $O(2N)$ calculations of lines,2,3.
}
\revised{
After we build the whole cluster tree, we randomly select one representative data point to calculate its objective value. After that, we use the continuous domination predicate mentioned in Section~\S\ref{ap} to rank all representative data points. The cluster which contains the representative data point in the first rank will be the best cluster.}

\revised{
For this algorithm,  Chen et al.~\cite{chen2018sampling} suggests to generate 10000 initial candidates, and then pruning that back to $\sqrt{n}$ candidates in each cluster.
We found that for generating test cases for simulation models, 256 initial random samples are enough to find good solutions (we made this conclusion after trying both 10000 and 256 initial \revised{set of} samples).}
\revised{
When utilizing 256 samples, we also found that Chen et al.'s recommendation of clustering down to  $\sqrt{n}$ was less useful that \textbf{split}ing down to ``enough'' (from Algorithm 1) being set to 4 and 16 and 32 is more relative to our intuitive. (we choose 4 as test suite size to check if algorithm can effectively generate small test suite, while we also check size of 16 and 32 to make sure if the best algorithm for generating small test suite can completely detect all mutants when test suite size increases).}


\section{Baseline Algorithms}\label{methodology}

For the purposes of comparisons, our GenClu system was compared against state-of-the-art systems that represent
\revised{the major themes in prior work}:
\bi
\item \revised{Matinnejad et al.'s 2018  OD evolutionary programming systems~\cite{matinnejad2018test};  see \S\ref{od}}.
\item \revised{Haq et al.'s 2022 SAMOTA surrogate model based optimization system~\cite{haq2022efficient}; see \S\ref{samota}}.
\ei
\revised{Note that, recalling 
\S\ref{epi}, we do not explore EPIcuRus since the nature of testing in that system is different to GenClu, OD, and SAMOTA.}

\subsection{OD Test Case Generation}\label{od}
Output Diversity (OD) test case generation approach is proposed by Matinnejad et al. on 2019~\cite{matinnejad2018test}. This approach based on meta-heuristic search to discover the maximum diversity in the simulation output signals~\cite{matinnejad2018test}. More specifically, OD firstly sets the initial number of signal pieces $P$ permitted in test inputs as 1, and tweak parameter $\sigma$ as 0.5. The initial random test suite is generated based on the input ranges and initial number of signal pieces $P$. After that, within the specific timeout value (we set same value as previous work, which is 600 seconds), OD algorithm repeatedly does following steps.
\begin{itemize}
    \item If it's the first iteration, then just skip this step. If not, it generates \textit{new test suite} by applying  the tweak algorithm on the \textit{recorded candidates} and current tweak parameter $\sigma$. \revised{Here the recorded candidates means test cases saved from previous iterations and the tweak parameter is used to control the scaler of tweak operation.}
    \item Simulate the test cases in the \textit{new test suite} to obtain the output signals.
    \item Calculate the coverage achieved by test cases in the \textit{new test suite}, and add it to the accumulative coverage. \revised{Here the coverage is the structural coverage in Simulink such as ``MC/DC'' and the accumulative coverage is the structural coverage achieved by the current test cases.}
    \item Compute the output diversity objective value $O$ for the \textit{new test suite}. If the objective value is higher than the recorded one, then replace the \textit{recorded test suite} by \textit{new test suite}, and record highest objective value.
    \item Check if accumulative coverage has reached a plateau at a value less than \%100. If reached, then increase the number of signal pieces $P$ by 1.
    \item If accumulative coverage is increased over initial coverage in current step, reduce tweak parameter $\sigma$ proportionally from $\sigma$\textit{-exploration} (0.5) to $\sigma$\textit{-exploitation} (0.01).
\end{itemize}

\subsection{SAMOTA} \label{samota}
\revised{ The SAMOTA algorithm   proposed by Haq et al. on 2022~\cite{haq2022efficient} is based on surrogate model optimization to optimize the objectives. More specifically, SAMOTA consists mainly of two search algorithms called ``global search'' and ``local search''. In the global search, the global surrogate models are trained on the basis of the test cases that exist in the database. With those global surrogate models, global search returns the best test case (ie, with the highest fitness score) for each uncovered objective~\cite{haq2022efficient}. After exploring the landscape of fitness functions in the global search, local search then exploits those promising regions returned from the global search. Firstly, the local search groups the candidates based on their fitness values using the HDBSCAN. Then for each cluster, the local surrogate model is trained by the candidates in that cluster. In such a scenario, the local search can explore all promising areas with a limited number of surrogate models~\cite{haq2022efficient}. As long as a better test case is established for an uncovered objective through a global and local search, the original test case for that uncovered objective will be replaced. The search will stop when certain iterations are achieved or when the test budget is exceeded. The original SAMOTA is designed for the binary classification task (which means the system needs to input the human-defined violation criteria). To replicate SAMOTA in our experiment, we replace that violation criteria with the objective values of a randomly generated candidate. Whenever a higher score is discovered during the search, that objective will be marked as covered objective.} 

\section{Experimental Setup}\label{experiment}

\subsection{Case Studies}\label{caseStudy}

For case studies for this work, we searched on (a) literature on testing \revised{simulation models} and (b) the official Simulink website ``MathWorks''. Our goal is to find case studies that satisfy the following requirements.
\begin{itemize}
    \item The model must be runnable in the model-in-the-loop level (which means it can be simulated without any hardware or sensors).
    \item The model must be developed in Simulink since all our experimental scripts are written in Matlab and rely on the simulations on Simulink.
    \item The simulation model must use discrete solvers. Continuous solvers will result in different length of output signals with same inputs, which we cannot apply mutation testing on them.
    \item The simulation model must have enough modules that can be mutated. We will introduce how we mutate the system in one of the following sections.
\end{itemize}

\begin{table}[!b]
    \centering
    \caption{Summary of case studies. In the table, we collect number of inputs and outputs (\# I/O), number of blocks (\# Blocks), number of original mutants (\# OM), number of filtered mutants (\# FM), and the mutant filtering percentage (Filtered \%). }
    \label{tab:case_study}
    \footnotesize
    \begin{tabular}{c|c|c|c|c|c}
        Project & \# I/O & \# Blocks & \# OM & \# FM & Filtered \% \\
        \hline
        Tiny & 3/1 & 20 & 33 & 11 & 33\% \\
    
        Two tanks & 11/7 & 449 & 190 & 15 & 8\% \\
       
        CC & 6/2 & 82 & 35 & 12 & 34\% \\
      
        CLC & 2/7 & 101 & 60 & 18 & 30\% \\
        
        CW & 15/4 & 273 & 88 & 30 & 34\%
    \end{tabular}
\end{table}

After filtered by above two criteria, we found 5 open-source case studies that can be used in our study:
\bi 
\item "Tiny" is a simple physical model~\cite{arrieta2019pareto}; 
\item "Two tanks" simulates the incoming and outgoing flows of two tanks~\cite{menghi2019generating};
\item "CW"    simulates the electrics and mechanics of four car windows~\cite{arrieta2019pareto};
\item "CC" simulates the cruise controller system of a car~\cite{arrieta2017search} 
\item "CLC" simulates a rotating clutch system. 
\ei
These systems are summarized  Table~\ref{tab:case_study}. As we can see, except Two tanks case study, the mutant filtering percentages (number of mutants after preprocessing / number of mutants generated manually) are all around 30\%-35\%.   Two tanks has many arithmetic operations,
which generated many   mutations. We will explain how we create those mutants manually and how we preprocess those mutants before they can be applied into our study in next subsection~\S\ref{mutation}

In two of these five case studies, inputs are combined with boolean values and numeric values. If boolean values are directly implemented into \textbf{split} algorithm \revised{described} in Section~\S\ref{clusterapproach}, the calculation of distance may result \revised{in an} error. To mitigate this problem, we treat all boolean values as a numeric value between 0 and 1 in \revised{the} \textbf{split} algorithm. Then we replace the normal input block to a switch block for each boolean input such that if the input value is less than 0.5, the output of the switch block will be 0, and otherwise, the output will be 1.

\subsection{Mutants Generation}\label{mutation}

The oracle problem, discussed above (in \S\ref{mutationtesting}), prevented us from assessing model output by asking a human.
Instead, we used the mutation testing approach introduced in S\ref{mutationtesting}. Recall
that mutation testing seeds faults into correct models to create mutants and check if test cases can ``kill'' these mutants.
A test case ``kills'' when any of the value in the output vector from mutant model is different to the corresponding value from original model.

In this study, we search the Simulink fault patterns in the literature~\cite{binh2012mutation, brillout2009mutation, yin2014research, zhan2005search, matinnejad2018test}.
We found mutations very similar to those
documented by   Matinnejad et al. in  2018~\cite{matinnejad2018test}. These Simulink fault patterns are listed in Table~\ref{tab:fault_pattern}. 

\begin{table}[b!]
    \centering
    \caption{Simulink Fault Patterns Collected in Literature~\cite{binh2012mutation, brillout2009mutation, yin2014research, zhan2005search}, and also listed in~\cite{matinnejad2018test}.}
    \label{tab:fault_pattern}
    \begin{tabular}{p{0.18\textwidth}p{0.25\textwidth}}
        Pattern & Illustration \\
        \hline
        Change of constant values & Change constant value $c$ to some other values if it is numeric. If it is boolean, negating it.\\
        \hline
        Change of arithmetic operators & Switch +/- or replace + to $\times$. \\
        \hline
        Change of relation operator & Switch $\geq$ to $<$ or $\leq$, and switch $\leq$ to $>$ or $\geq$. \\
        \hline
        Change of logical operator & Switch \textbf{AND}, \textbf{OR} and \textbf{XOR}; Remove \textbf{NOT} or add \textbf{NOT}. \\
        \hline
        Incorrect Connection & Switch the input lines of the ``Switch'' block. \\
        \hline
        Incorrect Signal Data Type & Switch the ``double'' data type to ``single'', or switch ``fixdt(0,8,3)'' data type to ``fixdt(0,8,2)''. \\
        \hline
        Wrong Initial conditions and delay values & Change the initial value in ``Integration'' and ``Unit Delay'' blocks.
    \end{tabular}
\end{table}

We manually seed the faults by using the above patterns. When we seed the faults, we consider all possibilities to mutant a single component. For example, if a component is an arithmetic operator which adds the first two inputs and subtracts the third input (++-), we consider all combinations of + and - (which means it can have 8 mutants with different orders of +/-) as well as the multiplication of all those inputs (which means this component can generate 9 mutants).

After we generate all the mutants for a single case study, we want to remove those mutants that (a) killed by all test cases, (b) cannot be detected by any test case, and (c) can be detected by same test cases with other mutants (e.g. for those mutants have same outputs on all test cases, we only keep one and discard rests). 

We ran mutation testing for these mutants on 200 random test cases and filtered out mutants with the above three rules. The number of original mutants, filtered mutants, and filtering percentage is shown in last three columns in Table~\ref{tab:case_study}. Even some case studies have small number of mutants, we find it is hard to detect all of them, which is a hard problem in test case generation.

\begin{table*}[]
    \centering
    \begin{tabular}{c|c||c|c|c|c|c|c||c}
        \multirow{2}{*}{project} & \multirow{2}{*}{algorithm} & \multicolumn{2}{c|}{test suite size - 4} & \multicolumn{2}{c|}{test suite size - 16} & \multicolumn{2}{c||}{test suite size - 32} & \multirow{2}{*}{wins} \\
        \cline{3-8}
         & & median MS & IQR & median MS & IQR & median MS & IQR & \\
        \hline
        \hline
        \multirow{4}{*}{Tiny} & Random & 0.73 & 0.55 & 0.82 & 0.09 & \cellcolor{gray!15}0.91 & \cellcolor{gray!15}0.09 & 1 \\
         & OD & \cellcolor{gray!15}0.82 & \cellcolor{gray!15}0.18 & \cellcolor{gray!15}1.00 & \cellcolor{gray!15}0.09 & \cellcolor{gray!15}0.91 & \cellcolor{gray!15}0.09 & \textbf{3} \\
         & SAMOTA & 0.73 & 0.18 & - & - & - & - & 0 \\
         & GenClu & \cellcolor{gray!15}0.82 & \cellcolor{gray!15}0.09 & \cellcolor{gray!15}0.91 & \cellcolor{gray!15}0.09 & \cellcolor{gray!15}0.91 & \cellcolor{gray!15}0.09 & \textbf{3} \\
        \hline
        \hline
        \multirow{4}{*}{CLC} & Random & \cellcolor{gray!15}0.94 & \cellcolor{gray!15}0.05 & \cellcolor{gray!15}1.00 & \cellcolor{gray!15}0.00 & \cellcolor{gray!15}1.00 & \cellcolor{gray!15}0.00 & \textbf{3} \\
         & OD & 0.78 & 0.05 & 0.67 & 0.00 & 0.67 & 0.00 & 0 \\
         & SAMOTA & - & - & 0.94 & 0.11 & - & - & 0 \\
         & GenClu & \cellcolor{gray!15}0.89 & \cellcolor{gray!15}0.11 & \cellcolor{gray!15}1.00 & \cellcolor{gray!15}0.06 & \cellcolor{gray!15}1.00 & \cellcolor{gray!15}0.00 & \textbf{3} \\
        \hline
        \hline
        \multirow{4}{*}{CC} & Random & 0.92 & 0.25 & \cellcolor{gray!15}1.00 & \cellcolor{gray!15}0.00 & \cellcolor{gray!15}1.00 & \cellcolor{gray!15}0.00 & 2 \\
         & OD & 0.96 & 0.12 & \cellcolor{gray!15}1.00 & \cellcolor{gray!15}0.00 & \cellcolor{gray!15}1.00 & \cellcolor{gray!15}0.00 & 2 \\
         & SAMOTA & \cellcolor{gray!15}1.00 & \cellcolor{gray!15}0.00 & - & - & - & - & \textbf{1} \\
         & GenClu & \cellcolor{gray!15}1.00 & \cellcolor{gray!15}0.08 & \cellcolor{gray!15}1.00 & \cellcolor{gray!15}0.00 & \cellcolor{gray!15}1.00 & \cellcolor{gray!15}0.00 & \textbf{3} \\
        \hline
        \hline
        \multirow{4}{*}{CW} & Random & \cellcolor{gray!15}0.97 & \cellcolor{gray!15}0.07 & \cellcolor{gray!15}1.00 & \cellcolor{gray!15}0.00 & \cellcolor{gray!15}1.00 & \cellcolor{gray!15}0.00 & \textbf{3} \\
         & OD & 0.83 & 0.16 & 0.76 & 0.05 & 0.82 & 0.01 & 0 \\
         & SAMOTA & - & - & 0.97 & 0.07 & - & - & 0 \\
         & GenClu & \cellcolor{gray!15}0.97 & \cellcolor{gray!15}0.07 & \cellcolor{gray!15}1.00 & \cellcolor{gray!15}0.00 & \cellcolor{gray!15}1.00 & \cellcolor{gray!15}0.00 & \textbf{3} \\
        \hline
        \hline
        \multirow{3}{*}{Twotanks} & Random & \cellcolor{gray!15}0.87 & \cellcolor{gray!15}0.26 & \cellcolor{gray!15}1.00 & \cellcolor{gray!15}0.07 & \cellcolor{gray!15}1.00 & \cellcolor{gray!15}0.00 & \textbf{3} \\
         & SAMOTA & - & - & \cellcolor{gray!15}1.00 & \cellcolor{gray!15}0.13 & - & - & \textbf{1} \\
         & GenClu & \cellcolor{gray!15}0.80 & \cellcolor{gray!15}0.40 & \cellcolor{gray!15}1.00 & \cellcolor{gray!15}0.27 & \cellcolor{gray!15}1.00 & \cellcolor{gray!15}0.00 & \textbf{3}
    \end{tabular}
    \caption{\revised{Mutation scores for 5 case studies. The comparison is conducted with 3 different test suite size - 4, 16, and 32. The approach succeed in small test suite (i,e, 4) means the test cases generated by that approach are effective and efficient. We also check test suite size of 16 and 32 to explore the convergence rate of performance in different approaches. \colorbox{gray!15}{Light gray} cells mark the winning approach based on the statistical testing and the last column summaries how much times the approach in the top rank in the statistical testing.}}
    \label{tab:result1}
\end{table*}

\subsection{Performance Criteria}\label{evaluationMetric}
To evaluate the performance of each test case selection technique, we utilize the mutation score metric. Mutation score is the percentage of number of mutants that are being detected among all mutants (e.g. if we detect 8 mutants among 10 mutants, the mutation score will be $8/10 = 0.8$). The larger mutation score is, the better test cases are generated.

\subsection{Statistical Analysis}\label{statisticalAnalysis}
For the generality, we repeat each approach 20 times. To compare the performance of different generation approaches across those 20 repeats, we applied Scott-Knott analysis which ranks the test case generation approaches by their median mutation scores.

More specifically, Scott-Knott recursively partitions the list of candidates ($c$) into two sub-lists ($c_1$ and $c_2$) which the expected mean value before and after the division should be maximized~\cite{emblem, xia2018hyperparameter, 9463120}:
\begin{equation}
    E(\Delta) = \frac{\textrm{len}(c_1) * |\overline{c_1}-\overline{c}| + \textrm{len}(c_2) * |\overline{c_2}-\overline{c}|}{\textrm{len}(c)}
\end{equation}
Hypothesis test is then applied to check if two sub-lists differ significantly by using the Cliff's delta procedure. The delta value is $Delta = (\#(x>y) - \#(x<y))/(\textrm{len}(c_1)*\textrm{len}(c_2))$ for $\forall x \in c_1$ and $\forall y \in c_2$. To explain that, Cliff's delta estimates the probability that a value in the sub-list $c_1$ is greater than a value in the sub-list $c_2$, minus the reverse probability~\cite{macbeth2011cliff}. Two sub-lists differ significantly if the delta is not a ``small'' effect ($Delta >= 0.147$)~\cite{hess2004robust}.

The reason we choose Scott-Knott because (a) it is fully non-parametric and (b) it reduces the number of potential errors in the statistical analysis since it only requires at most $O(log2(N))$ statistical tests for the $O(N^2)$ analysis.
Other researchers also advocate for the use of this test~\cite{gates78}   since it   overcomes a common limitation of alternative multiple-comparisons statistical tests (e.g., the Friedman test~\cite{friedman1937use}) where   treatments are assigned to multiple groups (making it hard for an experimenter to   distinguish the real groups  where the means should belong~\cite{Carmer1985PairwiseMC}.

\section{Results}\label{result}
In this section, \revised{we explore three RQs and answer the RQs based on our experimental results.}

\textbf{RQ1: Which approach is the most effectiveness one on generating small test suite on open-source \revised{simulation models}?} To answer RQ1, we run (a)~the random approach, (b)~OD, \revised{(c)~SAMOTA} and (d)~our proposed clustering approach GenClu 20 times with different random seed in each repeat. We then applied Scott-Knott analysis described in Section~\S\ref{statisticalAnalysis}. 
We say that  two algorithms differ significantly if
their mutation scores have different ranks within our Scott-Knott analysis. 

Table~\ref{tab:result1} shows the performance of different algorithms. \revised{Note that SAMOTA only has performance score in test suite size of 4 in Tiny and CC, and in test suite size of 16 in CLC, CW, and Twotanks. The reason is that SAMOTA is the surrogate model based approach which uses the surrogate models generated for each uncovered objective to optimize each goal. Thus the number of generated test cases by SAMOTA mostly depends on the number of objectives. As described in Section~\ref{background}, the number of objectives in our experiment is $4*$number of output signals. Thus, the replication of SAMOTA in this experiment can generate various number of test cases in different case studies. Moreover, OD algorithm is not applicable to Twotanks case study since all inputs in Twotanks are constant signals. Therefore, OD's tweak algorithm is not suitable for this case study.}

\begin{table*}[!t]
    \centering
    \begin{tabular}{c|c||c|c|c|c|c|c|c|c|c|c}
        \multirow{2}{*}{test suite size} & \multirow{2}{*}{algorithm} & \multicolumn{2}{c|}{Tiny} & \multicolumn{2}{c|}{CLC} & \multicolumn{2}{c|}{CC} & \multicolumn{2}{c|}{CW} & \multicolumn{2}{c}{Twotanks} \\
        \cline{3-12}
         & & runtime & slower & runtime & slower & runtime & slower & runtime & slower & runtime & slower \\
        \hline
        \hline
         \multirow{5}{*}{4} & Random & 151s & 1.0 & 293s & 1.0 & 458s & 1.0 & 2340s & 1.0 & 950s & 1.0 \\
         & OD & 12861s & 85.2 & 8323s & 28.4 & 11495s & 25.1 & 15381s & 6.6 & - & - \\
         & SAMOTA & 15500s & 102.6 & - & - & 63848s & 139.4 & - & - & - & - \\
         & GenClu & 317s & \cellcolor{gray!15}2.1 & 481s & \cellcolor{gray!15}1.6 & 881s & \cellcolor{gray!15}1.9 & 3000s & \cellcolor{gray!15}1.3 & 1418s & \cellcolor{gray!15}1.5 \\
        \hline
        \hline
        \multirow{5}{*}{16} & Random & 608s & 1.0 & 1207s & 1.0 & 2363s & 1.1 & 9024s & 1.0 & 3613s & 1.0 \\
         & OD & 16659s & 27.4 & 18242s & 14.4 & 21536s & 10.4 & 35525s & 3.9 & - & - \\
         & SAMOTA & - & - & 432530s & 358.4 & - & - & 91789s & 10.2 & 117110s & 33.1 \\
         & GenClu & 669s & \cellcolor{gray!15}1.1 & 1283s & \cellcolor{gray!15}1.1 & 2062s & \cellcolor{gray!15}1.0 & 10704s & \cellcolor{gray!15}1.2 & 3538s & \cellcolor{gray!15}1.0 \\
        \hline
        \hline
        \multirow{5}{*}{32} & Random & 3061s & 1.0 & 4415s & 1.0 & 9041s & 1.0 & 32954s & 1.0 & 12845s & 1.0 \\
         & OD & 20446s & 6.7 & 24253s & 5.5 & 119131s & 14.8 & 174183s & 5.3 & - & - \\
         & SAMOTA & - & - & - & - & - & - & - & - & - & - \\
         & GenClu & 4598s & \cellcolor{gray!15}1.5 & 5126s & \cellcolor{gray!15}1.2 & 8509s & \cellcolor{gray!15}1.1 & 43591s & \cellcolor{gray!15}1.3 & 13121s & \cellcolor{gray!15}1.0 \\
    \end{tabular}
    \caption{\revised{Runtime comparison for 5 case studies. \colorbox{gray!15}{Light gray} cells mark the runtime comparison for our proposed GenClu method. We say an approach is more efficient if its runtime is as close as the baseline Random method. As we can see, GenClu in only 1-2 times slower (i,e, in the worst case, 2.1 times slower) than Random while all other methods are numerous times to ten more times slower than Random.}}
    \label{tab:result2}
\end{table*}

As mentioned above, for the sake of completeness, we repeated our analysis for test suite sizes of 4, 16 and 32. Though 4 is a small test suite size, for our case studies, GenClu can get highest mutation score in all case studies with such small test suite size. For 16 and 32, we did not find an appreciable difference between them since most of the approaches get 1.00 mutation score across all studies. 

From Table~\ref{tab:result1}, we can see with test suite size of 4 (the \revised{left} part in Table~\ref{tab:result1}), GenClu \revised{has the highest performance} on all 5 case studies when random only wins on 3 and OD only wins on 1. \revised{SAMOTA is available in Tiny and CC in test suite size of 4, where SAMOTA gets lower performance in Tiny and similar performance in CC comparing to GenClu.} More specifically, with only 4 test cases in the test suite, GenClu can detect all faults in CC case study while the other three mostly miss few. In other four case studies, though GenClu cannot achieve 100\% mutation score, it still has the highest performance comparing to other approaches.

When test suite size is increased to 16 (the middle part in Table~\ref{tab:result1}), we can see that comparing to the size of 4, all approaches get similar or higher performance except OD algorithm. To explain that, we inspect the log information from simulation and find that OD runs much fewer iterations comparing to test suite size of 4 (e.g. 30+ iterations when size of 4 and only 8 to 10 iterations when size of 16) within the certain timeout parameter since the test suite size is increased a lot. This shows that OD found less informative candidates with fewer iterations, and thus get lower performance when test suite size is increased to 16. \revised{In test suite size of 16, SAMOTA only detects all mutants in Twotanks case study, and has lower performance in CLC and CW comparing to GenClu}. By looking at results of test suite size of 16, GenClu gets highest performance in all case studies again comparing to all other approaches.

Though we find that most of the performance in test suite size of 16 already convergent to 100\%, we also record the result of test suite size of 32 since the traditional sample size in each iteration in OD is 30, and 32 is the closest number to 30 which can \revised{be} divided by 256. As we expected, except OD, other approaches get either similar performance if its already convergent in size of 16 or improved performance if its not convergent in size of 16.

To summarize above, GenClu always stays in the most top rank whenever test suite size is 4, 16, or 32 across all case studies. Random does not performance bad on test suite size of 32, but lose on 1 or 2 case studies when test suite size is 16 and 4 correspondingly. OD does not performance well in test suite size of 4 and get even lower performance in test suite size of 16 and 32 in CLC and CW case studies due to less available iteration runs in the certain timeout setting. Though we suggest to increase the timeout limit in OD when test suite size is increased, it can make the overall runtime for OD much lengthy and exceed the time budget in testing. Overall, we answer RQ1 as:

\begin{tcolorbox}[boxsep=1pt,left=4pt,right=4pt,top=2pt,bottom=2pt]
    In {\bf all} projects, GenClu can always be found in the best rank. More specifically, \revised{GenClu is the only approach that has the highest mutation score in all 5 case studies when test suite size is 4}. When test suite size increases to 16, GenClu can dected 91\% faults in Tiny and all faults in the rest 4 case studies. When test suite size is 32, performance of GenClu and random is convergent in all case studies while OD still lose on few case studies. In summary, \revised{GenClu is the most effective approach in all different size of test suite.}
\end{tcolorbox}

\begin{figure*}
    \centering
    \includegraphics[width=1\textwidth]{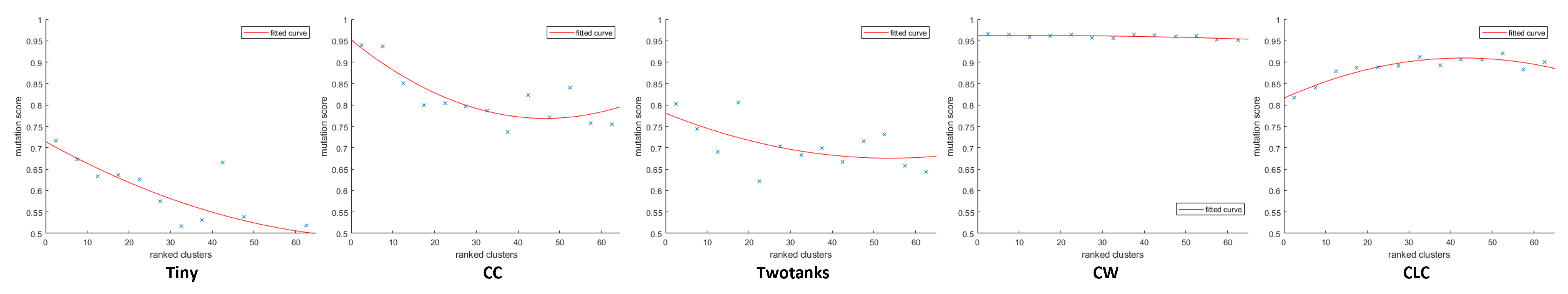}
    \caption{The scatter plots the brute-force mutation score for all the clusters ranked by the GenClu. The scatters from left to right on x-axis indicates the best rank to the worst rank, and y-axis is the mutation score. To avoid mass scatters, we plot the average mutation score for each five clusters, and plot the best fitting curve based on those scatters.}
    \label{fig:cluster_trend}
\end{figure*}

\textbf{RQ2: Which approach is the most efficient one on generating small test suite on open-source \revised{simulation models}?} To answer RQ2, we record the overall runtime (algorithm runtime plus simulation runtime plus mutation testing time) for each approach. To make comparison fair enough, we ran all three algorithms on the same 64-bit Windows 10 machine with a 4.2 GHz 8-core Intel Core i7 processor and 16 GB of RAM.

\revised{These runtimes can be seen in Table~\ref{tab:result2}. Further, under each case study, we summarize how many times the algorithm slower than the fastest one. As we observed, the baseline random method is the fastest one in most cases since it requires no information gain. Therefore, all coefficients in ``slower'' column will mostly be based on the runtime of random approach.}

As seen in Table~\ref{tab:result2}, \revised{GenClu can achieve   better performance than random method (as we showed in RQ1) and only requires 1 to 2 times longer duration than random method.} Moreover, comparing to the state-of-the-art method, GenClu can run 4-41 times faster than OD, \revised{and 40-300 times faster than SAMOTA}. 
As to why GenClu runs much faster than the other three:
\begin{itemize}
    \item OD tweaks the input signals through multiple iterations and calculates the output diversity in each iteration. It requires multiple simulations in each generation. 
    \item \revised{SAMOTA generates surrogate models for each uncovered objective, and optimize each objective based on surrogate models in multiple iteration. Thus, when number of objective goes up, the runtime of SAMOTA will increase exponentially.}
    \item GenClu, on the other hand,  can split random initial large samples into different clusters without any simulation, and find the best cluster with only one
    simulation from each group (since we split candidates by using the cosine rule, candidates in each cluster are close to each other). In such case, GenClu can find optimal solutions with few less simulations.
\end{itemize}
In summary, we answer RQ2 as follow
\begin{tcolorbox}[boxsep=1pt,left=4pt,right=4pt,top=2pt,bottom=2pt]
    In {\bf all} projects, \revised{GenClu can run multiple times faster than all other state-of-the-art methods (i,e, 4-41 times faster than OD, and 40-300 times faster than SAMOTA). More important, GenClu only 1-2 times slower than the baseline Random method which requires no information gain. Thus, we can conclude GenClu as an efficient test case generation method for \revised{simulation models}.}
\end{tcolorbox}

\revised{
\textbf{RQ3: Can future testing benefit from the tree structure generated by GenClu?} To answer RQ3, we first conduct an empirical validation to the tree structure generated by GenClu, and then implemented the mutation operator used in the differential evolution algorithm to mutate the test cases. 
}

\revised{First of all, we need to verify that the clusters ranked by the GenClu are informative. More specifically, the clusters with higher ranking will more likely to have higher mutation score, and the clusters with lower ranking will more likely to have lower mutation score. To achieve that, we firstly run the normal GenClu execution, which generates the tree structure and splits the samples into 64 clusters. The normal GenClu process will then rank those 64 clusters by the anti-pattern values of each randomly sampled representative from the clusters. To validate that such ranking system is informative, we then execute a brute force simulation which simulates all candidates for each cluster, and show the mutation score of each cluster in a scatter plot in Figure~\ref{fig:cluster_trend}. Please note that the x-axis shows the clusters ranked by the normal GenClu execution (which the clusters close to the original point have higher anti-pattern values and the clusters close to the end of x-axis have lower anti-pattern values). The y-axis shows the average mutation score by simulating all test cases in the corresponding cluster. To show the trend of mutation score from the good clusters to the bad clusters, we take the average for each five clusters and generate the best fitting curve by those average values. As we can see from Figure~\ref{fig:cluster_trend}, the slope of fitting curves in Tiny and CC are very large, which indicates that the clusters ranked by GenClu are very informative. Though the slope of fitting curves in CW and Twotanks is lower than the slope in Tiny and CC, it still can indicate that the first few clusters have better performance than the other clusters. The only exception is CLC that the slope of fitting curve is opposite to other case studies. This showed that in 4 out of 5 case studies, the clusters generated and ranked by GenClu are informative, and can lead the algorithm to generate the best test cases.}

\revised{After validated that the ranked clusters generated by GenClu are informative, we then asked if the cluster structure generated by GenClu can benefit the future testing. On the conclusions of the EPIcuRus (discussed above) is that some ranges of input data are more informative than other ranges~\cite{gaaloul2020mining}. Inspired by their study, our empirical validation results indicate that the first few clusters ranked by GenClu can explore the most informative region in the input space. However, the challenge here is that instead of producing the most informative ranges, GenClu only split the data into different clusters where the only information inside each cluster is data points in that cluster. To mitigate this challenge, we utilize the mutation operation in the differential evolution algorithm. Differential evolution is a method to optimize the problem by iteratively improving the candidates. Intelligently uses the differences between individuals realized in a simple and fast linear operator to mutate the candidates in each iteration and find the global optimal~\cite{feoktistov2006differential}. In GenClu structure, we adopt the mutation operator in differential evolution algorithm to mutate the candidates in the best few clusters to generate more informative test cases. More specifically, the differential evolution algorithm mutates the data points with the following formula:
\begin{equation}
    p_{new, i} = 
    \begin{cases}
        a_i + F * (b_i - c_i) & \text{if $r_i < CR$ or $i = R$} \\
        p_{old, i} & \text{O.W.}
    \end{cases}
\end{equation}
where 
\begin{itemize}
    \item $a, b, c$ are three different candidates in the population.
    \item $r_i$ is a uniform random number $\in (0, 1]$.
    \item $R$ is a random index $\in \{1, 2, \cdots, n\}$.
    \item $F$ and $CR$ are parameters of mutation operator. We pick $F = 0.9$ to mutate for a large scale and $CR = 1.0$ to make sure every element in the candidate is mutated.
\end{itemize}
}

\revised{
Table~\ref{tab:improvement} presents the average improvement of mutation score from all original test cases to the mutated test cases. As we can see, four case studies have improved score on mutation testing. Tiny and CC have higher improvement as we expected since the best fitting curves in Figure~\ref{fig:cluster_trend} have larger slope. CW and Twotanks have minor improvement with same reason.
\begin{table}[]
    \centering
    \begin{tabular}{c|c|c|c}
        Case Study & Original Score & Improvement & Improved Score \\
        \hline
        Tiny & 0.599 & +0.242 & 0.841 \\
        \hline
        CC & 0.808 & +0.072 & 0.880 \\
        \hline
        CW & 0.958 & +0.017 & 0.975 \\
        \hline
        Twotanks & 0.744 & +0.051 & 0.795 \\
        \hline
        CLC & 0.881 & -0.026 & 0.855
    \end{tabular}
    \caption{Average improvement from original randomly generated test cases to mutated test cases.}
    \label{tab:improvement}
\end{table}
}

\revised{In summary, we answer RQ3 as follow
\begin{tcolorbox}[boxsep=1pt,left=4pt,right=4pt,top=2pt,bottom=2pt]
    \revised{The cluster structure generated by GenClu is informative which the best few clusters can reflect more defective region in the input space. Moreover, by utilizing mutation operator in the differential evolution algorithm, we can mutate those clusters to produce new test cases with improved mutation score, which can benefit future testing.}
\end{tcolorbox}
}

\section{Threats to Validity}\label{threats_to_validity}
In this section, we will discuss the threats to validity issues raised by Feldt et al.~\cite{feldt2010validity}.

\textbf{Construct validity} mainly shows up in the parameter settings of algorithms. Different parameters may lead to quite different results. For example, Our replication experiment of EPIcuRus uses 30 iterations to generate the best input ranges. In different case studies, this number matters and may need to adjust properly. Moreover, in our proposed clustering approach, we randomly generate 256 initial \revised{set of} samples and cluster down to 4 and 16 and 32 for each bin. In different criteria or different case studies, these numbers need to be adjusted based on the requirements and purposes of study. The observation obtained from these parameters may differ if different parameters are used. We should consider hyper-parameter optimization tuning~\cite{tu2018while, tu2021mining} in future work to mitigate this threat.

\textbf{Conclusion validity} mainly \revised{relates} to the random variations of algorithms and the access to the real faults. To reduce the effect of this threat, we run all experiments 20 times for generality and use Scott-Knott statistical analysis to evaluate the performance through these 20 repeats. Moreover, since all case studies in our work are open-source projects, there is no public real faults on these software. To mitigate this threat, we better address the faulty patterns showed up in the past literature and collect them into a table (see Section~\S\ref{mutation}). We then generate mutants that cover all faulty patterns in that table. In this step, we make sure the mutants we generated are as close to the real faults as possible and use these mutants to better address the threat that caused by lacking access to the real faults.

\textbf{Internal validity} focuses on the correctness of the treatment caused the outcome. To reduce the effect of this threat, all experiments on different algorithms are running on the same case studies. Moreover, we compare three approaches in the same size of test suite. \revised{Another threat needs to be mentioned here is that our conclusion on the efficiency here is based on the execution of case studies applied in our experiment. Since the systems in this study may not need a lengthy simulation time, the runtime may vary on different case studies. However, our conclusion fully discusses the efficiency aspect regarding to the number of simulations needed for each algorithm. Therefore, our algorithm theoretically can reduce the simulation effort when applying to the more complex \revised{simulation models}.}

\textbf{External validity} concerns the application of this study in other problems. We utilize five open-source \revised{simulation models} to conduct our conclusions of this study. This may become a threat that five models cannot represent the society of \revised{simulation models}. However, we search these case studies from (a) past literature and (b) official website of Simulink and filter out the case studies which do not suitable in our work (Please see Section~\S\ref{caseStudy} for detailed information on selecting case studies). Also, \revised{simulation models} are hard to collect thus most of the previous studies only utilized 2-4 models (and most of them are private). Hence, we use the available resource to better address this threat in our work. Another threat may caused by applying algorithms in this study into other kinds of works. Both our replicated algorithm EPIcuRus and our proposed clustering algorithm need modification for different objective functions. We plan to apply these techniques into more \revised{simulation models} as future work to mitigate this threat. \revised{Moreover, some information like system coverage, system requirements, and environment uncertainty may unavailable to researchers, but they can be available to practitioners for the SUT that they are developing. Thus availability of those information can be a threat to this study}.





\section{Conclusion \& Future Work}\label{conclusion}
In this work, we (a) \revised{adopt} the idea of anti-pattern values on output signals to find potential faults on \revised{simulation models}, and (b) explore the application of {\em semi-supervised methods} in test case generation. The ``manifold assumption'', i.e. data with high dimension can be approximated to a lower dimensional manifold without much loss of signal~\cite{books/mit/06/CSZ2006}, is the key factor in the success of our  proposed approach. 

We evaluate GenClu on 5 \revised{simulation models} designed in Simulink and compare it to (a) a baseline random method, (b) the modified state-of-the-art method EPIcuRus, which includes the new test case generation algorithm inside its functionalities, (c) another state-of-the-art method OD test case generation, \revised{and (d) the most recent state-of-the-art online testing framework SAMOTA for DNN-enabled system.} By running experiments 20 repeats, we find {\em semi-supervised method based approach} GenClu can generate test cases with significant higher performance on mutation testing (win on all 5 case studies by passing the statistical test). By summarizing the results, we say that under less oracle condition, GenClu is a both effectiveness and efficiency method on generating test suite for detecting faults.

In future work, we plan to (a) collecting more \revised{simulation models} from other resources and replicate our study in other programming languages, (b) discover more potential anti-patterns which can reveal faults in the simulation models, (c) enhance the GenClu by adding some heuristics during the split algorithm, and (d) better handle the binary inputs by implementing the binary split algorithm in literature~\cite{chen2018sampling} into the current continuous split algorithm.

For other researchers exploring this area, or others in SE, we offer the following advice: we do not always have to explore all parts of the state space. Sometimes
(as shown here) a few ($\sqrt{n}$) peeks works as well as anything else (and does so much faster).


\bibliographystyle{IEEEtran}
\bibliography{ref}
\newpage
\begin{IEEEbiography}[{\includegraphics[width=1.05in,clip,keepaspectratio]{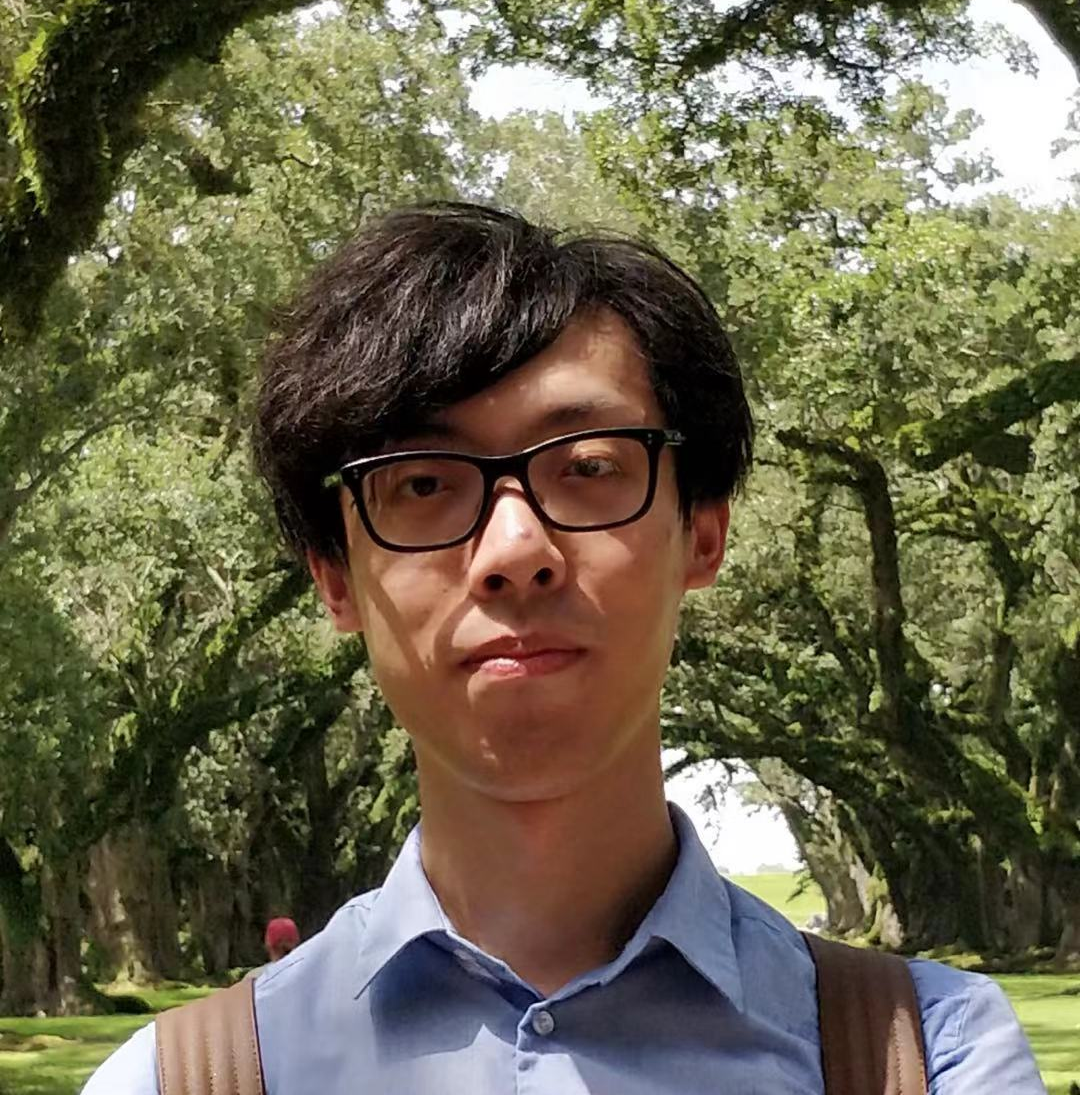}}]{Xiao Ling} is a fourth-year PhD student in Computer Science at NC State University. His research interests include automated software testing and machine learning for software engineering.
\end{IEEEbiography}

\begin{IEEEbiography}[{\includegraphics[width=1.05in,clip,keepaspectratio]{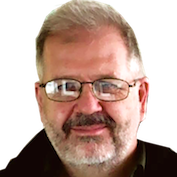}}]{Tim Menzies} (IEEE Fellow, Ph.D. UNSW, 1995)
is a Professor in computer science  at NC State University, USA,  
where he teaches software engineering,
automated software engineering,
and programming languages.
His research interests include software engineering (SE), data mining, artificial intelligence, and search-based SE, open access science. 
For more information,  please visit \url{http://timm.fyi}.
\end{IEEEbiography}
 
\end{document}